\documentclass[%
 aip,
 apl,%
 amsmath,amssymb,
 reprint,%
citeautoscript]{revtex4-1}

\usepackage{graphicx}
\usepackage{bm}

\usepackage{eso-pic}

\newcommand{\sss}{\scriptscriptstyle}
\newcommand{\sst}{\scriptstyle}

\newcommand{\stext}[1]{\sss \text{#1} \sst}

\newcommand{\rv}[1]{{#1}}

\makeatother

\begin{document}

\title{Electron effective mass in Al$_{0.72}$Ga$_{0.28}$N alloys determined by mid-infrared optical Hall effect}

\author{S.~Sch\"{o}che}
\email{schoeche@huskers.unl.edu} \affiliation{Department of Electrical Engineering and CNFM, University of Nebraska-Lincoln, Lincoln, 68588-0511, U.S.A.} \homepage{http://ellipsometry.unl.edu}
\author{P.~K\"{u}hne} \affiliation{Department of Electrical Engineering and CNFM, University of Nebraska-Lincoln, Lincoln, 68588-0511, U.S.A.}
\author{T.~Hofmann} \affiliation{Department of Electrical Engineering and CNFM, University of Nebraska-Lincoln, Lincoln, 68588-0511, U.S.A.}
\author{M.~Schubert} \affiliation{Department of Electrical Engineering and CNFM, University of Nebraska-Lincoln, Lincoln, 68588-0511, U.S.A.}
\author{D.~Nilsson} \affiliation{Department of Physics, Chemistry and Biology (IFM), Link\"oping University, Sweden}
\author{A.~Kakanakova-Georgieva} \affiliation{Department of Physics, Chemistry and Biology (IFM), Link\"oping University, Sweden}
\author{E.~Janz\'en} \affiliation{Department of Physics, Chemistry and Biology (IFM), Link\"oping University, Sweden}
\author{V.~Darakchieva} \affiliation{Department of Physics, Chemistry and Biology (IFM), Link\"oping University, Sweden}

\date{\today}

\begin{abstract}
The effective electron mass parameter in Si-doped Al$_{0.72}$Ga$_{0.28}$N is determined to be $m^\ast=(0.336\pm0.020)\,m_0$ from mid-infrared optical Hall effect measurements. No significant anisotropy of the effective electron mass parameter is found supporting theoretical predictions. Assuming a linear change of the effective electron mass with the Al content in AlGaN alloys and $m^\ast=0.232\,m_0$ for GaN, an average effective electron mass of $m^\ast=0.376\,m_0$ can be extrapolated for AlN. The analysis of mid-infrared spectroscopic ellipsometry measurements further confirms the two phonon mode behavior of the E$_1$(TO) and one phonon mode behavior of the A$_1$(LO) phonon mode in high-Al-content AlGaN alloys as seen in previous Raman scattering studies.

\end{abstract}

\date{\today}


\maketitle

Ternary alloys of AlGaN with high molar fractions of Al (high-Al-content AlGaN) are of high importance as the active layers in optical devices operated in the deep ultraviolet (UV) spectral range, including UV light emitting diodes and high-power UV laser diodes.  In order to predict accurate device functionality and optimize device design, knowledge of the basic electronic properties of the material, including the effective electron mass parameter, is essential. The free electron effective mass parameter in these materials has not yet been experimentally determined. 

The optical Hall effect (OHE), which comprises generalized spectroscopic ellipsometry measurements in combination with external magnetic fields, is a convenient and accurate method for characterization of free-charge carrier properties in semiconductor thin film heterostructures~\cite{Schubert2003a, Hofmann2003, Hofmann2006a, Hofmann2006, Hofmann2008a, Darakchieva2009, Schoeche2011}. This contact-less method provides access to the energy-distribution-averaged free-charge carrier parameters concentration, mobility, and effective mass~\cite{Schubert2003a,Schubert2004a,Hofmann2008a}. Prerequisite for  OHE application is the existence of sufficiently high ($\sim$${10}^{17}...10^{18}\,$cm$^{-3}$) free charge carrier concentration. Previous investigations on GaAs~\cite{Schubert2003a,Hofmann2007}, AlGaInP and BInGaAs alloys~\cite{Schubert2004a}, AlInP~\cite{Hofmann2008a}, ZnMnSe~\cite{Hofmann2006,Hofmann2006b}, InN~\cite{Hofmann2006,Hofmann2008,Darakchieva2009}, AlGaN/GaN high-electron mobility structures~\cite{Schoeche2011,Hofmann2012}, and graphene~\cite{Hofmann2011} showed that the OHE provides accurate values of the effective mass parameter that corroborate the results found, for example, from Shubnikov-de Haas measurements. Achieving reliable \emph{n}-type conductivity in high-Al-content AlGaN and AlN is very challenging due to the high ionization energy of the common donors (O and Si)~\cite{Neuschl2013}. Successful n-type doping of high-Al-content AlGaN and insights on the shallow/DX behavior of the Si donor has recently been reported~\cite{Kakanakova-Georgieva2013,Trinh2013}. 

For GaN, AlN, and AlGaN alloys with wurtzite crystal structure, an anisotropy of the effective electron mass parameter parallel and perpendicular to the $c$-axis is expected~\cite{ Rinke2008}. Mid-infrared spectroscopic ellipsometry (MIR-SE) in combination with electrical Hall effect measurements was previously applied by Kasic~\emph{et al.} in order to investigate the phonon mode parameters and anisotropic effective mass parameter in GaN~\cite{Kasic2000}. Slightly anisotropic effective electron mass parameters of $m^\ast_\parallel=(0.228\pm0.008)\,m_0$ parallel to the $c$-axis and $m^\ast_\perp=(0.237\pm0.006)\,m_0$ perpendicular to the $c$-axis were reported. For increasing Al-content in AlGaN alloys, an increasing value of the effective mass parameters is expected. Xu~\emph{et al.} demonstrated an increase of the transverse effective mass parameter (perpendicular to $c$-axis) from $m^\ast=0.27\,m_0$ to $m^\ast=0.30\,m_0$ with increasing Al-content for low molar fractions of Al between 0$\%$ and 52$\%$ by combining electrical Hall effect measurements with infrared reflectivity measurements~\cite{Xu2005}. So far, no clear experimental evidence is available indicating if this increase is proportional to the molar fraction of Al or if a second-order composition dependence has to be taken into account. Experimental data for the effective mass parameter in AlGaN alloys with molar fractions of Al larger than 52$\%$ are not available so far.

The most recent review of theoretical band parameters in the binary alloys GaN and AlN was given by Rinke~\emph{et al.}~\cite{Rinke2008} The predicted values for the effective electron mass in AlN are $m^\ast_\perp=0.33\,m_0$ perpendicular to the $c$-axis and $m^\ast_\parallel=0.32\,m_0$ parallel to the $c$-axis. These values corroborate the theoretical results by other authors~\cite{Vurgaftman2003, Carrier2005}. Combining these predicted values of the effective electron mass parameter with the experimental results for binary GaN and assuming a linear dependence of the effective mass parameter on the molar fraction of Al in ternary AlGaN alloys, only a small anisotropy of the effective mass is expected for high-Al-content AlGaN alloys.

In this work, we determine the effective electron mass parameter by using MIR-OHE and demonstrate high n-type conductivity in a high-quality, Si-doped Al$_{0.72}$Ga$_{0.28}$N thin epitaxial film. We further discuss the free-charge carrier and related phonon mode parameters in this high-Al-content AlGaN sample.

A Si-doped, epitaxial AlGaN film with a nominal thickness of 500\,nm was grown in a hot-wall metal-organic chemical vapor deposition (MOCVD) reactor on a on-axis semi-insulating 4H-SiC substrate~\cite{Kakanakova-Georgieva2009}. A graded buffer layer structure, consisting of an undoped AlN layer, a continuously graded AlGaN layer, and an undoped Al$_{0.72}$Ga$_{0.28}$N layer, is used for strain engineering in order to avoid cracking. Further details on growth conditions and structural analysis can be found in Ref.~\onlinecite{Kakanakova-Georgieva2013}. The Si concentration determined by SIMS and the net dopant concentration from C-V measurements are found to be  $9.6\times10^{18}$\,cm$^{-3}$. A custom-built Fourier transform-based MIR ellipsometer was used for the MIR-OHE measurements in the spectral range from 600\,cm$^{-1}$ to 1700\,cm$^{-1}$ with a resolution of 1\,cm$^{-1}$ \cite{Kuehne2013}. The MIR-OHE measurements were carried out at room temperature at an angle of incidence $\varPhi_{\stext{a}}=45^{\circ}$. The measurements were performed at $-7$\,T, $0$\,T and $+7$\,T, with the magnetic field oriented parallel to the incoming beam, resulting in a magnetic field strength $B_c=B/\sqrt{2}$ along the sample normal~\cite{Kuehne2013}.  The MIR-SE measurements have been carried out using a commercial Fourier transform-based MIR ellipsometer (J.A.~Woollam Co.~Inc.) in the spectral range from 300\,cm$^{-1}$ to 6000\,cm$^{-1}$ with a resolution of 2\,cm$^{-1}$.

The generalized ellipsometry formalism is employed for the optical investigations. The optical response of the sample is represented using the Mueller matrix formalism~\cite{Fujiwara_book2007}. In order to determine the phonon mode and free-charge carrier parameters, MIR-SE and MIR-OHE data were combined in a stratified layer model analysis using parameterized model dielectric functions. All model calculated data were matched simultaneously as closely as possible to the experimental MIR-SE and MIR-OHE data sets by varying relevant physical model parameters (best-model) \cite{Fujiwara_book2007}. The MIR spectral range dielectric functions of the sample constituents consist of contributions from optically active phonon modes $\varepsilon^{\text{L}}(\omega)$ and free-charge carrier excitations $\varepsilon^{\text{FC}}(\omega)$.  The dielectric functions of the nitride layers (wurtzite structure) and the 4H-SiC substrate substrate (hexagonal structure) are optically anisotropic (uniaxial). Functions $\varepsilon^{\text{L}}_j(\omega)$ are parameterized with Lorentzian lineshapes which account for transverse (TO) and longitudinal optic (LO) phonon frequencies, $\omega_{\stext{TO,}j}$ and $\omega_{\stext{LO,}j}$, respectively, for polarization $j$ =``$\parallel$'', ``$\perp$'' to the crystal $c$-axis \cite{SchubertIRSEBook_2004}:
\begin{equation}
\varepsilon^\mathrm{L}_j(\omega)=\varepsilon_{\infty,j}\prod\limits_{l}^{k}\frac{\omega^2+\mathrm{i}\gamma_{\stext{LO}l,j}\omega-\omega^2_{\stext{LO}l,j}}{\omega^2+\mathrm{i}\gamma_{\stext{TO}l,j}\omega-\omega^2_{\stext{TO}l,j}}.
\label{Eq:IR-model}
\end{equation}
$\varepsilon^{\text{FC}}(\omega)$ is parameterized using the classical Drude formalism \cite{Pidgeon80}. If magnetic fields are present the free-charge carrier Drude contribution is described by a tensor which allows for determination of the screened plasma frequency tensor $\bm{\omega}_{\stext{p}}$ and the cyclotron frequency tensor $\bm{\omega}_{\stext{c}}$\cite{Pidgeon80,Hofmann2008a}:
\begin{align}\label{DA}
&\bm{\varepsilon}^{\text{FC}}(\omega)  = \bm{\omega}_{\mbox{{\tiny
p}}}^{2}\\\nonumber
&\times\left[ -\omega^{2}\bm{I}-i\omega\bm{\gamma}+i\omega
\left(\begin{array}{ccc}
0 & b_{3} & -b_{2}\\
-b_{3} & 0 & b_{1}\\
b_{2} & -b_{1} & 0
\end{array}\right)
\bm{\omega}_{\mbox{\tiny c}} \right]^{-1}.
\end{align}
The free-charge carrier mobility tensor is given by $\bm{\mu}= q/(\bm{\gamma}\bm{m}^{\ast})$ where $\bm{m}^{\ast}$ denotes the effective mass tensor in units of the free electron mass $m_{0}$.  The plasma frequency tensor $\bm{\omega}_{\stext{p}}$ is related to the free-charge carrier density $N$ and the effective mass tensor $\bm{m}^{\ast}$ by $\bm{\omega}_{\stext{p}}^2=Nq^2/(\varepsilon_\infty\tilde{\varepsilon}_0 \bm{m}^{\ast}m_0)$, where $q$ denotes the charge, $\tilde{\varepsilon}_0$ is the vacuum permittivity, and $\varepsilon_{\infty}$ denotes the high frequency dielectric constant. The cyclotron frequency tensor is defined as $\bm{\omega}_{\mbox{{\tiny c}}}=qB/(m_{0})\bm{m}^{\ast-1}$. The external magnetic field is given by $\bm{B}=B(b_{1}, b_{2}, b_{3})$ with $|\bm{B}|=B$.

\begin{figure} \centering\includegraphics[keepaspectratio=true,trim= 0 2 0 2, clip, width=6.8cm]{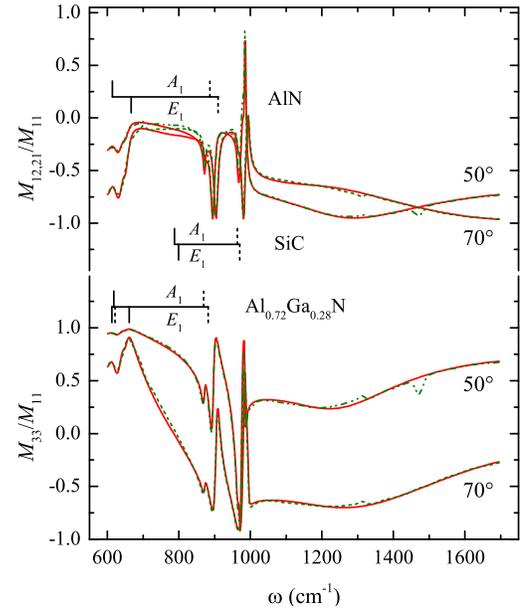}
\caption{Experimental (dotted lines) and best-model calculated (solid lines) $M_{12}$, $M_{21}$, and $M_{33}$ spectra for the 4H-SiC substrate/AlN buffer layer/graded AlGaN layer/undoped Al$_{0.72}$Ga$_{0.28}$N layer/Si-doped Al$_{0.72}$Ga$_{0.28}$N sample structure in the spectral range from 600\,cm$^{-1}$ to 1700\,cm$^{-1}$ for two angles of incidence of  $\varPhi_{\stext{a}}=50^{\circ}$ and $\varPhi_{\stext{a}}=70^{\circ}$. The frequencies of the IR-active phonon modes of the AlGaN layer, the AlN buffer, and the SiC substrate are indicated by brackets (solid: TO; dotted: LO). } \label{fig:IRSE}
\end{figure}

Figure~\ref{fig:IRSE} depicts the normalized MIR-SE $M_{12}$, $M_{21}$, and $M_{33}$ spectra of the heterostructure obtained at angles of incidence of $\varPhi_{\stext{a}}=50^{\circ}$ and $\varPhi_{\stext{a}}=70^{\circ}$. The experimental (broken lines) and best-model calculated (solid lines) data are found to be in very good agreement. The spectra are dominated by the typical features of group-III nitride layers on SiC substrates in the range of the reststrahlen bands of the sample constituents which are indicated by brackets in Fig.~\ref{fig:IRSE}~\cite{Kasic03}. The presence of a large number of free-charge carriers in the Si-doped Al$_{0.72}$Ga$_{0.28}$N layer is indicated in the $M_{12}$ and $M_{21}$ spectra by the presence of the peak structure at 1000\,cm$^{-1}$ which can be attributed to the well-known LO-phonon plasmon coupling effect in highly doped semiconductors~\cite{Kasic03}. 

For best-match model calculation, a model consisting of 4H-SiC substrate/AlN buffer layer/graded AlGaN layer/undoped Al$_{0.72}$Ga$_{0.28}$N layer/Si-doped Al$_{0.72}$Ga$_{0.28}$N layer was implemented with the dielectric functions of each layer modeled according to Eqn.~(\ref{Eq:IR-model}) and~(\ref{DA}).  The phonon mode frequencies and broadening parameters for the dielectric functions of the 4H-SiC substrate and the AlN were determined from MIR-SE measurements on a bare substrate and a single AlN layer grown on SiC under similar conditions as the sample presented in this investigation. These parameters were not further varied during the model analysis of the sample shown in Fig.~\ref{fig:IRSE}.
The best-model  phonon mode frequency and broadening parameters for the SiC substrate are $\omega_{\stext{TO,}\perp}=(798.0\pm0.2)$~cm$^{-1}$, $\omega_{\rv{\stext{LO,}\perp}}=(970.5\pm0.1)$~cm$^{-1}$,
$\gamma_{\perp}=(3.1\pm0.1)$~cm$^{-1}$,
$\omega_{\stext{TO,}\parallel}=788$~cm$^{-1}$,
$\omega_{\stext{LO,}\parallel}=(963.6\pm0.1)$~cm$^{-1}$, and
$\gamma_{\parallel}=(4.2\pm0.1)$~cm$^{-1}$ and are in good agreement with literature values~\cite{Tiwald1999}. 
The best-model  phonon mode frequency and broadening parameters for the AlN buffer layer are $\omega_{\stext{TO,}\perp}=(666.4\pm0.1)$~cm$^{-1}$, $\omega_{\rv{\stext{LO,}\perp}}=(909.9\pm0.3)$~cm$^{-1}$,
$\gamma_{\perp}=(3.2\pm0.2)$~cm$^{-1}$,
$\omega_{\stext{TO,}\parallel}=611$~cm$^{-1}$,
$\omega_{\stext{LO,}\parallel}=(886.4\pm0.1)$~cm$^{-1}$, and
$\gamma_{\parallel}=(9.2\pm0.1)$~cm$^{-1}$, also in good agreement with literature values~\cite{Darakchieva2004a}.
The thin graded AlGaN layer was accounted for by introducing a linearly graded layer (10 slices) for which the phonon mode parameters at the bottom and top of the layer were fixed to the parameters of the adjecent AlN and Al$_{0.72}$Ga$_{0.28}$N, respectively. Identical phonon mode parameters were used for the undoped and Si-doped Al$_{0.72}$Ga$_{0.28}$N layer. For polarization perpendicular to the $c$-axis, we observe a two-mode behavior with best-match model parameters $\omega_{\stext{TO,}\perp}^{\stext{AlN}}=(661.3\pm0.2)$~cm$^{-1}$, $\omega_{\stext{LO,}\perp}=(882.1\pm0.7)$~cm$^{-1}$, $\gamma_{\stext{TO/LO,}\perp}=(13.2\pm0.4)$~cm$^{-1}$, $\omega_{\stext{TO,}\perp}^{\stext{GaN}}=(611.5\pm0.5)$~cm$^{-1}$, $\omega_{\stext{LO,}\perp}=(621.3\pm0.5)$~cm$^{-1}$, $\gamma_{\stext{TO,}\perp}^{\stext{GaN}}=(23\pm1)$~cm$^{-1}$, and $\gamma_{\stext{LO,}\perp}=(28\pm1)$~cm$^{-1}$.    For polarization parallel to the $c$-axis, a single LO phonon mode frequency $\omega_{\stext{LO,}\parallel}=(869.1\pm0.2)$~cm$^{-1}$ is determined with $\gamma_{\stext{TO/LO,}\parallel}=(6.3\pm0.4)$~cm$^{-1}$. The TO phonon mode parameters for the polarization parallel to the $c$-axis are not accessible for c-plane oriented samples~\cite{SchubertIRSEBook_2004}. A constant value of $\omega_{\stext{TO,}\parallel}=611$~cm$^{-1}$ is used~\cite{Darakchieva2004a}. The determined phonon mode parameters are similar to those reported in Ref.~\onlinecite{Davydov2002} for thick AlGaN layers of comparable composition grown without buffer layers on Si-(111) substrates and analyzed by Raman spectroscopy. Our results confirm the two phonon mode behavior of the E$_1$(TO) mode and the one mode behavior of the A$_1$(LO) mode reported in Ref.~\onlinecite{Davydov2002}.

\begin{figure}[tb] \centering\includegraphics[keepaspectratio=true,trim=0 2 0 0, clip, width=7cm]{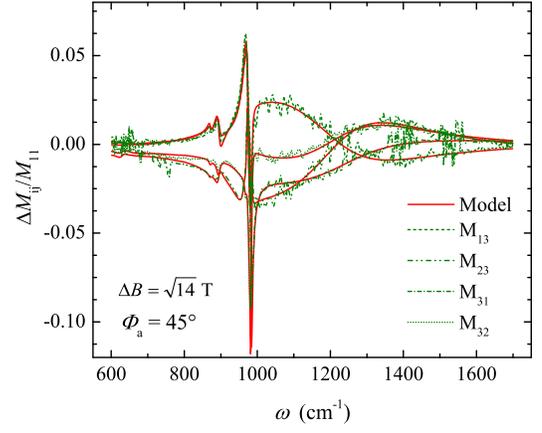}
\caption{Experimental (broken lines) and best-model calculated (solid lines) Mueller matrix difference spectra [$\Delta M_{ij}=M_{ij}(B=+7.0~$T$)-M_{ij}(B=-7~$T)] for the 4H-SiC substrate/AlN buffer layer/graded AlGaN layer/undoped Al$_{0.72}$Ga$_{0.28}$N layer/Si-doped Al$_{0.72}$Ga$_{0.28}$N sample structure in the spectral range from 600\,cm$^{-1}$ to 1700\,cm$^{-1}$ obtained at an angle of incidence $\varPhi_{\stext{a}}=45^{\circ}$ for magnetic field orientation along the incoming light beam.}\label{fig:OHE}
\end{figure}

The MIR-OHE spectra are presented in Fig.~\ref{fig:OHE}. The off-diagonal block Mueller-matrix elements $M_{13}$, $M_{23}$, $M_{31}$ and $M_{32}$ depict the magnet-field induced birefringence and render the optical Hall effect. The lattice term of the dielectric function $\varepsilon^{\text{L}}_j(\omega)$ is symmetric under magnetic field inversion while the free-charge carrier term $\varepsilon^{\text{FC}}(\omega)$ is antisymmetric. Therefore, the non-zero values seen in the differences of $M_{13}$, $M_{23}$, $M_{31}$ and $M_{32}$ for magnetic fields of $B=+7.0$\,T and $B=-7.0$\,T in Fig.~\ref{fig:OHE} are solely caused by the free-charge carrier related magneto-optical birefringence~\cite{Hofmann2008a}. This magneto-optical birefringence is strongest close to the LO-phonon plasmon coupled modes which is seen as the prominent features in Fig.~\ref{fig:OHE}~\cite{Hofmann2008a}. The spectral position of these features is mainly determined by the free-charge carrier concentration while the shape is influenced by the free-charge carrier mobility and effective mass parameters. The free-charge carrier related birefringence vanishes outside the range of the reststrahlen bands. Note, that only the five frequency-independent parameters free-charge carrier concentration $N$, effective masses $m^\ast_\parallel$ and $m^\ast_\perp$, and mobilities $\mu_\parallel$ and $\mu_\perp$ are sufficient to achieve the excellent match between experimental (broken line) and model data (solid lines) over the whole spectral range shown in Fig.~\ref{fig:OHE}.

The best-match model data in Fig.~\ref{fig:OHE} was calculated by applying the same model as for the MIR-SE data set during the simultaneous analysis of all data sets. The model analysis reveals $n$-type conduction in the Si-doped Al$_{0.72}$Ga$_{0.28}$N layer with a high volume free-charge carrier concentration of $N=(1.1\pm0.2)\times10^{19}$~cm$^{-3}$. This value is in excellent agreement with the net dopant concentration as determined by the C-V measurements.  No significant anisotropy of the effective mass parameter is found for this free-charge carrier concentration. The determined effective mass parameters are $m^\ast_\perp=(0.334\pm0.010)\,m_0$ perpendicular to the $c$-axis and $m^\ast_\parallel=(0.338\pm0.025)\,m_0$ parallel to the $c$-axis. The sensitivity of our technique for the effective mass parameter for polarization parallel to the $c$-axis $m^\ast_\parallel$ is slightly reduced compared to the sensitivity for $m^\ast_\perp$. However, from our data analysis, it can be estimated that the anisotropy of the effective electron mass parameter may not be higher than 0.03. Theoretical investigations of the effective mass for AlN do not predict a significant anisotropy of the effective mass at the $\Gamma$-point~\cite{Rinke2008}. Our result for the sample of relatively high Al-content investigated here supports the theoretical predictions and is also in agreement with reports of vanishing effective mass anisotropy for similar electron concentrations in InN and GaN~\cite{Hofmann2008}. Slightly different mobility parameters of $\mu_\perp=(39\pm1)$\,cm$^2$/Vs and $\mu_\parallel=(32\pm1)$\,cm$^2$/Vs are determined for the directions perpendicular and parallel to the c-axis, respectively. \rv{Anisotropic mobility parameters have been previously observed in GaN~\cite{Kasic2000} and InN~\cite{Hofmann2008} and could be related to different distributions of extended defects and impurities in directions parallel and perpendicular to the c-axis.} We further allowed for free-charge carriers in the SiC substrate and the AlN buffer layer, but no free-charge carrier contributions from these layers were revealed during the model analysis. 

Under the assumption of a linear increase of the effective mass parameter in AlGaN alloys with increasing Al-content and assuming an average effective mass value of $m^\ast=0.232\,m_0$ for electrons in GaN~\cite{Kasic2000}, an average effective electron mass value of $m^\ast=0.376\,m_0$ can be extrapolated for AlN. The only experimental estimate of the effective electron mass parameter in AlN was given by Silveira~\emph{et al.}~\cite{Silveira2004} By analyzing cathodoluminescence measurements and using effective hole mass parameters in AlN as derived from the Mg acceptor binding energies in AlN~\cite{Nam2003} a range for the effective electron mass parameter of 0.28\,$m_0$ to 0.45\,$m_0$ was estimated~\cite{Silveira2004}. Our estimated value for the effective electron mass in AlN fits well into this range. Kim~\emph{et al.} studied the free-charge carrier related LO-phonon plasmon coupled mode shift in a n-type Al$_{0.67}$Ga$_{0.33}$N alloy by means of Raman scattering~\cite{Kim2011a}. In order to model the line shape of the Raman data, the authors assumed a value for the effective electron mass parameter as determined from linear interpolation between the values for binary GaN and AlN. Good match between model and experimental data was achieved by using the theoretical value of $m^{\ast}=0.32\,m_0$ for AlN as given by Rinke~\emph{et al.}~\cite{Rinke2008} Our extrapolated value for the effective mass parameter for AlN is slightly larger then theoretically predicted value Rinke~\emph{et al.} However, it is known that \emph{ab initio} density-functional theory calculations used in Ref.~\onlinecite{Rinke2008} tend to give underestimated values of electronic parameters. 

In summary, we have determined an effective electron mass parameter in Al$_{0.72}$Ga$_{0.28}$N of 0.336\,$m_0$ by combining optical Hall effect and spectroscopic ellipsometry measurements in the MIR spectral range. No significant anisotropy is found for the effective electron mass parameter supporting theoretical predictions. Assuming a linear change of the effective electron mass with the Al content in AlGaN alloys and an effective electron mass parameter in GaN of $m^\ast=0.232\,m_0$, an average effective electron mass of $m^\ast=0.376\,m_0$ is extrapolated for AlN. We further confirm the two phonon mode behavior of the E$_1$(TO) and one phonon mode behavior of the A$_1$(LO) phonon mode in high-Al-content AlGaN alloys as seen in previous Raman scattering studies. 

The authors acknowledge financial support from the National Science Foundation under awards MRSEC DMR-0820521, MRI DMR-0922937, DMR-0907475, and EPS-1004094, by the Swedish Research Council (VR) under grant No.2010-3848, and the Swedish Governmental Agency for Innovation Systems (VINNOVA) under the VINNMER International Qualification program, grant No. 2011-03486. A.K.G. acknowledges support from the Link\"{o}ping Linnaeus Initiative for Novel Functionalized Materials (VR) and VINNOVA.


\end{document}